# Tunable spin textures in polar antiferromagnetic hybrid organic-inorganic perovskites by electric and magnetic fields

Feng Lou[1,2,†], Teng Gu[1,2,†], Junyi Ji[1,2], Junsheng Feng[3], Hongjun Xiang[1,2,4,*], and Alessandro Stroppa[5,*]

[1]*Key Laboratory of Computational Physical Sciences (Ministry of Education), State Key Laboratory of Surface Physics, and Department of Physics, Fudan University, Shanghai 200433, China.*

[2]*Collaborative Innovation Center of Advanced Microstructures, Nanjing 210093, China.*

[3]*School of Physics and Materials Engineering, Hefei Normal University, Hefei 230601, China*

[4]*Shanghai Qizhi Institution, Shanghai 200433, China*

[5]*CNR-SPIN, c/o Department of Physical and Chemical Sciences, University of L'Aquila, 67100 Coppito (AQ), L'Aquila, Italy*

*Email: hxiang@fudan.edu.cn, alessandro.stroppa@spin.cnr.it*
† These authors contribute equally to this work.

**Abstract**

The hybrid organic-inorganic perovskites (HOIPs) have attracted much attention for their potential applications as novel optoelectronic devices. Remarkably, the Rashba band-splitting, together with specific spin orientations in *k*-space (*i.e.* spin texture), has been found to be relevant for the optoelectronic performances. In this work, by using first-principles calculations and symmetry analyses, we study the electric polarization, magnetism, and spin texture properties of the antiferromagnetic (AFM) HOIP ferroelectric TMCM-MnCl$_3$ (TMCM=(CH$_3$)$_3$NCH$_2$Cl$^+$, trimethylchloromethyl ammonium). This recently synthesized compound is a prototype of order-disorder and displacement-type ferroelectric with a large piezoelectric response, high ferroelectric transition temperature, and excellent photoluminescence properties [You *et al.*, Science 357, 306 (2017)]. The most interesting result is that the inversion symmetry breaking coupled to the spin-orbit coupling gives rise to a Rashba-like band-splitting and a related robust persistent spin texture (PST) and/or typical spiral spin texture, which can be manipulated by tuning the ferroelectric or, surprisingly, also by the AFM magnetic order parameter. The tunability of spin texture upon switching of AFM order parameter is largely unexplored and our findings not only provide a platform to understand the physics of AFM spin texture but also support the AFM HOIP ferroelectrics as a promising class of optoelectronic materials.

## Introduction

The past few years witnessed the extremely rapid development of hybrid organic-inorganic perovskites (HOIPs), which have been shown to be promising optoelectronic material [1-5]. HOIP materials have several commom features, including the classical $ABX_3$ perovskite architecture and the presence of organic cation that occupy the A-site. As for the B-site, it can be occupied not only by main group elements, but also by transition metal atoms such as Mn and Fe, thus introducing magnetic degrees of freedom into the compound. As for the X-site, it is usually the halogen element. The HOIP materials have some advantages and in particular, the exceptionally long carrier lifetimes make them very attractive for optoelectronic devices, such as light absorbers and light-emitting diodes [6-11].

To further enhance the optoelectronic performances of HOIP materials, intense research has been directed to explain the microscopic origin of the long lifetimes [9-11]. Recently, the presence of Rashba band-splitting has been suggested to be connected with the carrier lifetimes and to improve their optoelectronic performances [12-15]. When lacking spatial inversion symmetry, the spin-orbit coupling (SOC) effect leads to an effective momentum-dependent magnetic field $\vec{\Omega}(\vec{k})$ acting on the spin $\vec{\sigma}$, and the effective SOC Hamiltonian can be written as $H_{SO} = \vec{\Omega}(\vec{k}) \cdot \vec{\sigma}$ [16-17]. In this case, the SOC will split the spin degeneracy with specific spin orientations (*i.e.* spin texture) in the momentum *k*-space, as was firstly demonstrated by Rashba [18] and Dresselhaus [19]. The spin texture can often be manipulated and even reversed by switching the electric polarization under external electric field, leading to an all-electric and non-volatile control of spin state [20-24]. Rashba effects were mainly discussed in non-magnetic lead halide perovskites [9-15, 24-28] or non-magnetic ferroelectric semiconductors [20-23, 29-33]. However, to the best of our knowledge, there are no studies on the spin texture in AFM HOIP ferroelectrics. Furthermore, antiferromagnets are very appealing for spintronic applications due to their superior properties, since they produce no stray fields and display intrinsic ultrafast spin dynamics [34-36]. In the last few years, intense theoretical and experimental research has shown that it is possible to realize electric-field control of magnetism in multiferroic materials [37-39]. The couplings between polarization, magnetism, and spin textures are still largely unexplored but they could have important applications in magneto-optoelectronic devices. Indeed, some recent reviews have pointed out intriguing spin-optotronic properties in HOIP materials [40-41]. Therefore, it is interesting to study whether there exists unusual Rashba effects in AFM HOIP ferroelectrics and how to manipulate the spin textures.

In this work, we start by considering the compound of TMCM-MnCl$_3$ (TMCM=(CH$_3$)$_3$NCH$_2$Cl$^+$, trimethylchloromethyl ammonium). In 2017, You *et al.* reported that TMCM-MnCl$_3$ is a ferroelectric material which shows an excellent piezoelectric response ($d_{33}$=185 pC/N) that is close to that of inorganic piezoelectrics such as BaTiO$_3$ ($d_{33}$=190 pC/N) and a high transition temperature $T_c$ of 406 K. Besides, TMCM-MnCl$_3$ displays excellent photoluminescence properties, with a near-unity photoluminescence emission efficiency [42]. In our study, we discuss the interplay among ferroelectric and magnetic orderings, and spin textures by using density-functional theory (DFT). We show that TMCM-MnCl$_3$ is a prototype of order-disorder and displacement-type ferroelectric whose polarization can be greatly modified by the halogen atom substitutions. The most important result is that a Rashba-like effect in the band structure leads to robust unidirectional persistent spin texture (PST) and/or spiral spin texture [31, 43]. The spin textures have been predicted to support an extraordinarily long spin lifetime which is promising for optoelectronic devices [44-45]. By tuning the ferroelectric or, surprisingly, the antiferromagnetic order parameter, we find that the spin texture can be modified significantly. Our results indicate that not only the electric but also the magnetic field can effectively be used to manipulate the spin textures even in AFM but polar HOIP materials such as TMCM-MnCl$_3$. Our results suggest that AFM HOIP ferroelectric is an interesting class of materials which deserves further study.

**Results**

**Structural properties.** At high temperature, TMCM-MnCl$_3$ adopts a paraelectric phase with centrosymmetric space group *P6$_3$/mmc* with disordered organic cations. However, as the temperature decreases, TMCM-MnCl$_3$ undergoes an order-disorder phase transition at around 406 K and crystallizes into the polar *Cc* phase [42]. In order to find the experimental ground-state structure, we rotate the organic cations randomly to consider different configurations, and then optimize these structures to calculate their total energies. In our work, we generate about 100 random structures, and find that the *Cc* phase indeed has the lowest total energy, in agreement with the experimental result. It has a monoclinic conventional cell with the distortion angle between *a* and *c* axis about 95°. Our optimized lattice constants are *a*=9.371, *b*=15.548, and

$c$=6.521 which are consistent with the experimental report of $a$=9.478, $b$=15.741, and $c$=6.577 (Å). As shown in Fig. 1, TMCM-MnCl$_3$ contains four organic cations and four Mn ions in the conventional cell. Their crystal packing is similar to BaNiO$_3$-like perovskite. The Mn ions form the inorganic chains along [001] direction with the ligand ions of Cl, while the organic cations are inserted between these inorganic chains. One can see that the freezing of polar organic cations can give rise to the ferroelectric polarization along approximately the [10$\bar{1}$] direction. From crystallographic analysis, it has six different polar axes with 12 possible orientations of polarization. This multiaxial characteristic is certainly interesting for fundamental research and practical applications of HOIP ferroelectrics [46].

**Ferroelectric and magnetic properties.** In order to study ferroelectric properties, we apply the modern theory of electric polarization [47-48]. The details of DFT calculations are described in Section Methods. To simulate the AFE-FE transition, we fix two organic cations and rotate the other two cations by introducing an interpolating parameter λ [see Note S1 and Fig. S1 of the Supplemental Material (SM)] [49-51]. Note that this dimensionless parameter λ is not the usual linear interpolation for atomic positions but it defines the correlated rotation of cations as well as the displacement of the MnCl$_3$ framework. Therefore, it represents the normalized amplitude of the roto-displacive path. Since the transition path is artificially assigned to act as a computational tool, the polarization with λ=1 has a real physical meaning. Here we define our convention for the coordinates as the *x* (*y*) axis being along *a* (*b*) axis, respectively. As for the *z* axis, it is vertical to the *x*-*y* plane and it has an angle about 5° with *c* axis. In TMCM-MnCl$_3$ system, the polarization is evaluated to be 6.12 μC/cm$^2$ approximately along the [10$\bar{1}$] direction (P$_x$ is about 4.18 μC/cm$^2$ and P$_z$ is about -4.48 μC/cm$^2$), which is in rather good agreement with the experimental value of P$_z$ ~ 4.00 μC/cm$^2$ [42]. To shed light into the microscopic mechanism of FE polarization, we perform mode decomposition [52] with respect to the reference centric phase by considering the different functional units, *i.e.* organic cations and framework. This approach, called functional mode analysis, has been already used for the analysis of FE polarization in hybrid compounds. Here functional refers to functional units in the HIOPs, *i.e.*, organic cations and framework. It is useful to disentangle the different contributions to the total polarization by considering the role played by the different functional units. We find that the polarization contains two main contributions, one is from the organic cation about 4.87 μC/cm$^2$, and the other part comes from the distortion of the inorganic framework which is about 1.15 μC/cm$^2$. The first contribution can be associated to the ordering of the organic cations, while the second one can be related to a significant displacement-type contribution. Therefore, TMCM-MnCl$_3$ is a prototype as order-disorder and displacement-type ferroelectric.

The halogen atoms and H atoms can form a complex *hydrogen bonding* network with the organic cations, which mainly determine the relative orientations of the organic cations with respect to the framework. Therefore, it may be useful to study how the halogen substitutions may influence the ferroelectric polarization. Indeed, the halogen atoms have similar chemical properties, but they differ in electronegativity, which, in turn, will effectively change the electric polarization through hydrogen bond network that is responsible for the complex cations and framework interaction. By changing the halogen atoms in the inorganic framework and/or organic cations, we find that the polarization can be significantly modified (see Fig. S2-S3 of SM).

As for the magnetic ground state, we performed collinear calculations showing that TMCM-MnCl$_3$ has strong AFM interaction within the inorganic MnCl$_3$ chains. This can be understood in terms of Goodenough-Kanamori (GK) rule which predicts a strong AFM super-exchange interaction between two half-filled $Mn^{2+}$ ($3d^5$) ions [53-54]. However, the interchain interaction between the inorganic MnCl$_3$ chains is weak AFM since the distance between neighboring chains is large (more than 9 Å). The energy of different magnetic configurations is shown in Fig. S4 of SM. The G-type AFM state is the ground state with AFM intrachain and interchain couplings. To accurately evaluate the spin coupling parameters, we adopt four-state method [55-56]. The effective spin exchange $J$ for the intrachain Mn-Mn pair is computed to be 12 meV, while the interchain interaction is about 0.1 meV. When considering the spin-orbit coupling effect, the non-collinear calculations show that the local spin moments tend to be perpendicular to the MnCl$_3$ chains and the magnetic anisotropy energy (MAE) is about 0.03 meV/Mn. We note that there is no relevant canting of spins in TMCM-MnCl$_3$ system, *i.e.* we have a collinear AFM HOIPs compound. Considering that TMCM-MnCl$_3$ has weak interchain interaction, one can apply external fields (*e.g.,* magnetic field) to switch the spins along one direction, *i.e.,* C-type AFM state (intrachain AFM coupling and interchain FM coupling). Therefore, considering the tunable ferroelectric and magnetic states, the TMCM-MnCl$_3$ system provides an ideal platform to investigate the interplay between ferroelectric ordering, magnetic ordering, and spin textures.

**Band structure properties.** We investigate the electronic properties of valence band maximum (VBM) and conduction band minimum (CBM) by calculating the band structures with/without SOC (see Fig. 2a-2d). Here the conventional cell containing four organic cations and four Mn ions (see Fig. 1). When considering SOC, the spin moment is set to be along *y* axis. The band structures of G-type $AFM_y$ and C-type $AFM_y$ states are shown in Fig. 2b and 2d, respectively. To help understand the spin textures discussed in the following paragraphs, we choose a specific symmetric *k*-path containing $k_b$ and $k_{ac}$, which is perpendicular to the polarization (see Fig. 2e and 2f) [23, 31-32, 57]. Here $k_b$ and $k_{ac}$ denote the *k* path from Γ (0,0,0) to Y (0,0.5,0) and Q (0.5,0,0.5), respectively. In order to simplify the illustration of Brillouin zone, we simplify the crystal lattice from slightly monoclinic to orthorhombic (see Fig. 2f). For G-type AFM state (see Fig. 2a), our calculations show that the VBM and CBM are located at Γ point, and the partial density of states (DOS) show that the valence band edge contains contributions from Mn-3*d* and Cl-2*p* orbitals, whereas the conduction band edge is mainly composed of Mn-3*d* orbitals (see Fig. S5). Due to the symmetry (see below for detailed analysis), all eigenstates are at least two-fold degenerate (i.e., spin-up and spin-down states). When taking SOC into account, the Rashba-Dresselhaus effect removes the spin degeneracy into singlets along the symmetry path but it still keeps two-fold degeneracy at Γ point (see Fig. 2b). Interestingly, for the C-type AFM state, the doublet at Γ point splits into two singlets with a sizable spin splitting at VBM about 0.027 eV after inclusion of SOC (see Fig. 2d).

To understand the band degeneracy at Γ point, we perform the symmetry analysis by considering Kramers degeneracy. Considering a Hamiltonian $\hat{H}$ with an eigenvector $\psi$ and a real eigenvalue $\lambda$ such as $\hat{H}\psi = \lambda\psi$. Let $\psi' = \hat{A}\psi$, where $\hat{A}$ commutes with $\hat{H}$. It's easy to write: $\hat{H}\psi' = \hat{H}\hat{A}\psi = \hat{A}\hat{H}\psi = \lambda\hat{A}\psi = \lambda\psi'$. Hence, both $\psi$ and $\psi'$ are eigenvectors of $\hat{H}$ with the same eigenvalue $\lambda$. One can prove that $\psi$ and $\psi'$ are orthogonal to each other if $\hat{A}$ is anti-unitary and $\hat{A}^2\psi = -\psi$, since $\langle\psi,\hat{A}\psi\rangle = \langle\hat{A}\psi,\hat{A}^2\psi\rangle^* = -\langle\hat{A}\psi,\psi\rangle^* = -\langle\psi,\hat{A}\psi\rangle$ and thus $\langle\psi,\hat{A}\psi\rangle = 0$. Due to the orthogonality, $\psi$ is degenerate with $\psi' = \hat{A}\psi$. Therefore, if $\hat{A}$ commutes with $\hat{H}$ and $\hat{A}^2 = -1$, the band structure can be double degenerate. In our

TMCM-MnCl$_3$ system, the G-type AFM$_y$ state has the magnetic symmetry of $\hat{M} = \{\{E|0\}, \{\hat{m}_{ac}|\frac{a+b+c}{2}\}, \hat{T}\{\hat{m}_{ac}|\frac{c}{2}\}, \hat{T}\{E|\frac{a+b}{2}\}\}$, where $E$ is identity operator, $\hat{T}$ is time-reversal operator, $\hat{m}_{ac}$ is the mirror symmetry operator followed by lattice translation. We find that the operator $\hat{A}_b = \hat{T}\{E|\frac{a+b}{2}\}$ is anti-unitary and commutes with the Hamiltonian at Γ point (here $\hat{A}_b$ plays a similar role as $\hat{T}$ in the time-reversal invariant case considered by the Kramers degeneracy). The band structure of G-type AFM$_y$ state is shown in Fig. 2b and we use the subscript of $\hat{A}$ (i.e., b) to index the band structure. Using the properties of half-spin system at Γ point, we can identify $\hat{A}_b^2 = \hat{T}^2 = -1$, leading to two-fold degeneracy at Γ point with SOC effect. As for C-type AFM$_y$ state, it has the magnetic symmetry of $\hat{M} = \{\{E|0\}, \{E|\frac{a+b}{2}\}, \hat{T}\{\hat{m}_{ac}|\frac{a+b+c}{2}\}, \hat{T}\{\hat{m}_{ac}|\frac{c}{2}\}\}$. Different from G-type AFM$_y$ state, we cannot find such an anti-unitary symmetry operator $\hat{A}_d$ to construct Kramers pair, since $\left(\hat{T}\{m_{ac}|\tau_c\}\right)^2 \psi = \hat{T}^2 m_{ac}^2 \psi = -1 \cdot (-1) \cdot \psi = \psi$, where $\tau_c = \frac{a+b+c}{2}$ or $\frac{c}{2}$. Hence, the energy bands of C-type AFM$_y$ with SOC are all singlet as shown in Fig. 2d. When turning off SOC, spin is independent from the spatial degrees of freedom and pure spin rotation $\hat{U}$ can be introduced to explain the energy band degeneracy [58]. $\hat{U}$ can reverse the spin but it is unitary and keeps the momentum invariant. For collinear AFM system without SOC, the wave function $\phi$ can be chosen to have a definite $S_z$ value (1/2 for up-spin or -1/2 for down-spin), thus $\hat{U}\phi$ and $\phi$ are orthogonal and form the Kramers pair. Without SOC, G-type AFM state has the symmetry of $\hat{U}\{E|\frac{a+b}{2}\}$. It commutes with the Hamiltonian for all wave vectors and lead to two-fold degeneracy in the whole BZ, including path Q − Γ − Y as shown in Fig. 2a. As for the C-type AFM state, the two-fold degeneracy along Q − Γ and Γ − Y (see Fig. 2c) can be ascribed to different symmetry mechanism. The wave vectors in Q − Γ and Γ − Y respect the symmetry of $\hat{T}\{m_{ac}|\frac{c}{2}\}$ and $\hat{U}\{m_{ac}|\frac{c}{2}\}$, respectively. Note that without SOC, one can get $m_{ac}^2 \psi = \psi$, and hence we can have $\hat{A}_c = \hat{T}\{m_{ac}|\frac{c}{2}\}$, $\hat{A}_c^2 \psi = \hat{T}^2 m_{ac}^2 \psi = -1 \cdot 1 \cdot \psi = -\psi$. Therefore, both $\hat{T}\{m_{ac}|\frac{c}{2}\}$ and $\hat{U}\{m_{ac}|\frac{c}{2}\}$ can form Kramers pair and the corresponding band structure is double degenerate. Apart from these symmetry arguments, we can also apply systematic group theory analysis based on the co-representation of the magnetic point group to understand the spin

degeneracy at Γ point (see Note S2 of SM). These two methods give same results. Therefore, to summary our discussion, the different symmetry operations in G-type AFM$_y$ state and C-type AFM$_y$ state can lead to different band degeneracy at Γ point.

**Manipulation of spin textures.** Knowing the spin degeneracy at Γ point, we can consider the spin texture around this point in the Brillouin zone. Considering that TMCM-MnCl$_3$ displays two long range ordering, *i.e.*, ferroelectric and AFM orderings, it is interesting to see how the spin textures behave under the interplay of these two order parameters. Recently, the electric-field control of spin textures has been shown in non-magnetic ferroelectric GeTe thin film [20-23]. Here, as we will show below, the spin textures in TMCM-MnCl$_3$ can be tuned by switching not only ferroelectric ordering but also by switching the magnetic order parameter in an AFM polar HOIP. This represents a new degree of freedom to play with in the spin-texture tuning, which has been very little studied in the literature. The SOC splits the band structure into two branches, which exhibit similar spin textures but with opposite helicity or orientation. Here, we will focus on the inner branches near Γ point, while the spin textures of the outer branches are illustrated in Fig. S7-S17 of SM. In order to simplify the visualization, we project the spin textures on a specific plane which is perpendicular to the polarization (see Fig. 2f) [23, 31-32, 57].

In the following, we discuss the spin textures in G-type AFM state. We pay attention to the spin texture at CBM, since the spin value at VBM is small due to the weak band splitting. It is useful to introduce the AFM order parameter defined as $\mathbf{L} = \sum_i \mathbf{S}^i - \sum_j \mathbf{S}^j$, where $\mathbf{S}^i$ ($\mathbf{S}^j$) is the spin moment along positive (negative) axis, respectively. We use the subscript of L to define different AFM state. For example, $\mathbf{L_G} \sim y$ indicates the G-type AFM configuration along *y* direction. And we use $\mathbf{L_G} \sim -y$ to indicate the operation that flip the spin from *y* to -*y* direction. The polarization (*P*) is along [10$\bar{1}$] direction while -*P* is along [$\bar{1}$01] direction. As we can see in Fig. 3a, it shows robust persistent spin texture (PST) at CBM. The spin is unidirectional and parallel ($k_{ac}<0$) or antiparallel ($k_{ac}>0$) to $k_b$ direction (*i.e.*, vertical to the mirror reflection). One can understand the spin texture by considering the magnetic symmetry. The G-type AFM$_y$ state has the magnetic space symmetry of $\widehat{M} = \{\{E|0\}, \{\widehat{m}_{ac}|\frac{a+b+c}{2}\}, \widehat{T}\{\widehat{m}_{ac}|\frac{c}{2}\}, \widehat{T}\{E|\frac{a+b}{2}\}\}$, which can be labeled with $\widehat{M} = \{\widehat{m}, \widehat{T}\widehat{m}, \widehat{T}\}$. Thus, one can have the following constraints on spin texture: $\mathbf{S}(k) = \hat{m}\mathbf{S}(\hat{m}k)$, $\mathbf{S}(k) = \hat{T}\hat{m}\mathbf{S}(\hat{T}\hat{m}k)$, and $\mathbf{S}(k) = \hat{T}\mathbf{S}(\hat{T}k)$. We note that the persistent spin textures occupy a substantial scale of Brillouin zone. It spans more than 0.04 Å$^{-1}$ around the Γ point, while for comparison the reciprocal wave vector of $k_b$ is π/b = 0.20 Å$^{-1}$ which corresponds to the length of symmetry path from (0,0,0) to (0,0.5,0). In this large area, the spin configurations remain nearly unidirectional which is favorable to support the long spin lifetime of carrier promising for optoelectronic applications [31, 43-44]. Our results suggest that TMCM-MnCl$_3$ is a rare example of Rashba-AFM HOIP ferroelectric with robust PST.

Here we discuss the interplay between ferroelectric ordering, magnetic ordering, and spin texture. In Fig. 3b, we fix the magnetic ordering but reverse the ferroelectric polarization from $P$ to $-P$. One can see that the PST is reversed with the switching of the polarization. The spin transformation rule under space inversion operator $\hat{I}$ is shown in Fig. 3e. When considering the spin configurations at the points $\vec{k}$ to $-\vec{k}$ related by the space inversion operator $\hat{I}$ (i.e. reversal of polarization), the spin orientations remain unchanged since $\hat{I}\mathbf{S}_\mathbf{P}(k) = \mathbf{S}_{-\mathbf{P}}(-k) = \mathbf{S}_\mathbf{P}(k)$, where $\mathbf{P}$ is the ferroelectric order parameter, $\mathbf{S}_{-\mathbf{P}}$ is the new spin after the space inversion. In G-type AFM$_y$ state, the spin at $-k$ point has opposite orientation compared with the spin at $k$ point, i.e., $\mathbf{S}_\mathbf{P}(-k) = -\mathbf{S}_\mathbf{P}(k)$, while after the space inversion (switching of polarization), one can get $\mathbf{S}_{-\mathbf{P}}(-k) = \mathbf{S}_\mathbf{P}(k)$. Therefore, we can reverse the PST by switching the ferroelectric ordering. As for switching the magnetic ordering, the time reversal operation $\hat{T}$ will reverse the magnetic ordering parameter $\mathbf{L}$, and spin texture changes according to $\hat{T}\mathbf{S}_\mathbf{L}(k) = \mathbf{S}_{-\mathbf{L}}(-k) = -\mathbf{S}_\mathbf{L}(k)$, where $\mathbf{S}_{-\mathbf{L}}$ is the new spin after the time reversal. In Fig. 3c, we fix the ferroelectric order, and then we flip magnetic ordering from $\mathbf{L_G} \sim y$ to $\mathbf{L_G} \sim -y$. We find the PST remains unchanged since $\mathbf{S_L}(-k) = -\mathbf{S_L}(k)$ and $\mathbf{S_{-L}}(-k) = -\mathbf{S_L}(k)$. Apart from the flip of magnetic ordering from $\mathbf{L}$ to $-\mathbf{L}$, one can also switch the magnetic ordering to other directions. In Fig. 3d, we rotate the magnetic ordering from $\mathbf{L_G} \sim y$ to $\mathbf{L_G} \sim x$. Surprisingly, there is PST at not only CBM (see Fig. 3d) but also VBM in comparison with the small spin value at VBM in $\mathbf{L_G} \sim y$ case (see Note S3 of SM). The G-type AFM$_x$ state has the magnetic symmetry of $\hat{M} = \{\{E|0\}, \{\hat{m}_{ac}|\frac{c}{2}\}, \hat{T}\{\hat{m}_{ac}|\frac{a+b+c}{2}\}, \hat{T}\{E|\frac{c}{2}\}\}$. The rotation part keeps the same with G-type AFM$_y$ state but the translational part of glide plane changes. We note that the magnetic symmetry is conserved as long as the magnetic ordering lies within the *ac* plane, *i.e.*, the mirror

$\hat{m}_{ac}$. According to the symmetry analysis, the operator $\hat{A} = \hat{T}\{E|\frac{c}{2}\}$ commutes with the Hamiltonian accompanied with $\hat{A}^2 = \hat{T}^2 = -1$ at Γ point, leading to two-fold degeneracy at Γ point with SOC effect (see Fig. S6 of SM). Our further calculations demonstrate that when we rotate the magnetic ordering within the *ac* plane, the PST can be switched along the magnetic ordering (see Note S3 of SM). It is an interesting result that we can continuously rotate PST by switching the magnetic ordering. Now we can draw the conclusion that the G-type AFM state shows the robust PST around Γ point and the PST can be manipulated not only by switching the polarization but also by switching the magnetic ordering. This is certainly a new result since so far, the switching of spin-texture chirality has been linked only to the switching of ferroelectric polarization while here we point out the active role of the switching of AFM order parameter.

Now we discuss the spin textures in C-type AFM state. TMCM-MnCl$_3$ has weak interchain interaction and one may apply external fields (e.g., magnetic field) to switch the spins along one direction, *i.e.* C-type AFM state. We find the PST along $k_{ac}$ direction, although there is a deviation from the unidirectional spin orientation when moving far away from the Γ point. We note that the PST in G-type AFM$_y$ state is along $k_b$ direction. This phenomenon shows that one can switch the PST by realizing different magnetic state. As for CBM, it exhibits spiral spin textures with clockwise helicity. This 2-dimensional vector field is identical to the characteristic Rashba-like spin texture [23, 32, 57]. The C-type AFM$_y$ state has the magnetic symmetry of $\widehat{M} = \{\{E|0\}, \{E|\frac{a+b}{2}\}, \hat{T}\{\hat{m}_{ac}|\frac{a+b+c}{2}\}, \hat{T}\{\hat{m}_{ac}|\frac{c}{2}\}\}$, which can be labeled with $\widehat{M} = \{\hat{T}\hat{m}\}$. Correspondingly, the doublet state at Γ point is lifted into singlets by SOC with a sizable band spin splitting at VBM about 0.027 eV (see Fig. 2d). According to the $\hat{T}\hat{m}$ symmetry, the PST and spiral spin texture can be understood with $\mathbf{S}(k) = \hat{T}\hat{m}\mathbf{S}(\hat{T}\hat{m}k)$. At Γ point, there is no spin component along $k_b$ direction since $\hat{T}\hat{m}k=k$ and $\mathbf{S}(\Gamma) = \hat{T}\hat{m}\mathbf{S}(\Gamma)$. It is interesting that VBM and CBM have same magnetic symmetry but show PST and spiral spin texture, respectively. To the best of our knowledge, it is the first case about the coexistence of PST and spiral spin texture in the same compound.

In Fig. 4c and 4d, we fix the magnetic ordering but flip the ferroelectric polarization from *P* to -*P*. The spin textures are switched according to $\hat{I}\mathbf{S}_\mathbf{P}(k) = \mathbf{S}_{-\mathbf{P}}(-k) = \mathbf{S}_\mathbf{P}(k)$. For the VBM of C-type AFM$_y$ state, the spin at $-k$ and $k$ point have same orientation, i.e., $\mathbf{S}_\mathbf{P}(-k) = \mathbf{S}_\mathbf{P}(k)$, while after the space inversion, one can get $\mathbf{S_{-P}}(-k) = \mathbf{S_P}(k)$. Therefore, when switching the ferroelectric polarization, the PST maintains the same spin orientation (see Fig. 4c). However, as for the CBM, the spin at $-k$ and $k$ point have opposite orientation, i.e., $\mathbf{S_P}(-k) = -\mathbf{S_P}(k)$. After the space inversion, it can be $\mathbf{S_{-P}}(-k) = \mathbf{S_P}(k)$. Thus, the helicity of spiral spin texture is reversed (see Fig. 4d). It is interesting about the different tunability of VBM and CBM under same external field. It is also important to note that this compound has been recently synthesized and the switching of polarization has been realized with a well-defined *P-E* loop [42], therefore we expect that the manipulation of spin textures by the external electric field could be easily verified by experiments. In Fig. 4e and 4f, we fix the ferroelectric order but flip the AFM ordering from $\mathbf{L_C} \sim \mathbf{y}$ to $-\mathbf{y}$ to see the variation of spin texture. We find the PST of VBM (see Fig. 4e) is reversed whereas the spiral spin texture of CBM (see Fig. 4f) maintains the same helicity. The variation of spin texture is consistent with the rule of $\hat{T}\mathbf{S_L}(k) = \mathbf{S_{-L}}(-k) = -\mathbf{S_L}(k)$. For VBM, before switching **L**, $\mathbf{S_L}(-k) = \mathbf{S_L}(k)$. After switching **L**, $\mathbf{S_{-L}}(-k) = -\mathbf{S_L}(k)$. However, for CBM, before switching **L**, $\mathbf{S_L}(-k) = -\mathbf{S_L}(k)$. After switching **L**, $\mathbf{S_{-L}}(-k) = -\mathbf{S_L}(k)$. The variation is totally different from the spin texture tunability upon the change of ferroelectric polarization. Our results demonstrate that one can manipulate the spin textures by switching the AFM order parameter but independently from the electric degrees of freedoms. Furthermore, to the best of our knowledge, we present a unique case in the literature, where there is coexistence of PST and spiral spin texture in the same material.

We also investigate the spin textures with other magnetic configurations (see Fig. S7-S17 of SM). By manipulating the magnetic order parameter with different orientation and different magnetic state, the corresponding spin-texture will change accordingly and it is the origin of magneto-crystalline anisotropy [59]. This property is dual of the spin-texture electric-anisotropy first discussed in the HOIP material $(NH_2CHNH_2)SnI_3$ [60] where it has been shown that the spin-texture topology is modified significantly upon variations of the direction of the electric polarization. The sensitivity of the topology of spin-texture to variation/switching of the magnetic order parameter could have far reaching consequences in AFM spintronics, since this property could be exploited in AFM memory elements: the change in spin-texture topology of relevant electronic bands should be detectable in terms of magneto-optical Kerr effect, as already shown in the metal-organic framework material $[C(NH_2)_3]Cr[(HCOO)_3]$ [61]. Further study is in progress to verify these properties. Our results clearly suggest that one could manipulate the spin texture via tuning the magnetic ordering at different levels: by fixing the magnetic configurations but changing the **L** orientation in space, or by changing the different realization of **L**. It has been shown that AFM materials can be manipulated by applying magnetic fields [35, 62-63]. The magnetic moments can be appreciably rotated in a quasi-static manner within the Stoner-Wohlfarth model [64]. In this picture, the ordered magnetic state is preserved when the magnetization is reversed and a spin-flop field can rotate the magnetic moments by 90° [65]. Besides, the AFM state could be reoriented by optical excitation [66-67], exchange bias [68-69], strain [70-71], and other different approaches [35, 63]. We note that the manipulation of ferroelectric polarization and magnetic configuration was realized in the classical multiferroic material $TbMnO_3$ [38] and $BiFeO_3$ [39]. Therefore, the ferroelectric and magnetic orderings in a polar AFM HOIP system could be tuned and the spin texture can be manipulated at the same time, thus leading to interesting magneto-optoelectronic applications.

Our study shows the possibility of tuning spin textures by electric and magnetic fields in AFM HOIP ferroelectrics and enhancing its optoelectronic performance, although there remain some challenges such as the wide bandgap and low magnetic ordering temperatures. In our TMCM-$MnCl_3$ system, the bandgap is calculated to be 3.75 eV which is larger than the ideal bandgap suitable for optoelectronic applications. It is reported that TMCM-$MnCl_3$ displays excellent photoluminescence properties with a near-unity photoluminescence emission efficiency [42], thus our DFT calculations may overestimate the bandgap. R.G. Xiong *et. al* proposed that TMCM-$MnCl_3$ can be further engineered through element substitution and molecular design so as to optimize for a desired physical properties, as shown by bandgap engineering [72-73]. Taking the characteristic HOIP material MA$PbI_3$ as an example, the band gap can be easily tuned from 1.2 to 3.0 eV by engineering chemical composition [73]. The magnetic ordering temperature can be improved as well. In our TMCM-$MnCl_3$ system, the Néel temperature is low due to the weak interchain interaction, which can be ascribed to the large organic cation. The Néel temperature could be improved with smaller organic cation. Furthermore, the substitution on B-site magnetic ions can enhance the magnetic ordering temperature. For example, it is reported that the HOIP material $((CH_3)_4P)FeBr_4$ exhibits coupled dielectric and magnetic phase transitions above room temperature [74]. It is important to note that our study puts forward the concept that one can manipulate the spin texture by applying magnetic fields. The external magnetic field can stabilize the AFM ordering and raise the Néel temperature. Therefore, we hope to stimulate the search of high temperature AFM HOIP ferroelectrics in the future.

**Conclusions**

In this work, we propose the manipulations of spin textures in the AFM HOIP ferroelectric TMCM-MnCl$_3$. By using first-principles calculations, we identify a Rashba-like splitting in the band structure. The symmetry analyses based on magnetic space group are used to explain the band degeneracy. We find robust PST in G-type AFM state and it can be effectively manipulated by switching not only polarization but also magnetic ordering. We also find the coexistence of PST and typical spiral spin texture, depending of the relevant band electronic states, in C-type AFM state. To the best of our knowledge, this is the first case of coexistence of PST and spiral spin texture in the same compound. By manipulating the ferroelectric, and, interestingly, the magnetic order parameter, the spin texture can be modified significantly. Our work introduces new directions in the field of spin-texture manipulation by external fields, which goes beyond the usual electric-field control of Rashba effect in non-magnetic materials. Considering that TMCM-MnCl$_3$ belongs to the important class of HOIPs, which is relevant to optoelectronic research, we expect that, this study could suggest new magneto-optoelectronic properties in HOIPs. Since the switching of polarization in TMCM-MnCl$_3$ has been experimentally demonstrated [42], we hope to stimulate new experiments to verify manipulations of spin-textures in TMCM-MnCl$_3$ by electric and/or magnetic fields. We expect that AFM HOIP ferroelectrics have the potential to improve the optoelectronic performance and give a new strategy to design new multifunctional materials.

**Methods**

**DFT calculations.** Our first-principles calculations are performed within density functional theory (DFT). The interactions of valence electrons and ions is treated with the projector augmented wave (PAW) method [75] as implemented in the Vienna *ab*-initio simulation package (VASP) [76]. The exchange-correlation potential is described by the Perdew-Burke-Ernzerhof (PBE) functional [77]. The plane wave cutoff energy is fixed to be 550 eV, and all atomic positions are optimized until each component of the atomic force is smaller than 0.01eV/Å. The $4 \times 2 \times 4$ k-point mesh is used for the Brillouin integration. The electric polarization is calculated by using the Berry phase method [48, 78]. In this approach, we first define a centrosymmetric reference phase which shows an antiferroelectric (AFE) alignment of dipoles in the unit cell and then we continuously rotate and translate the organic cations to reach the ferroelectric (FE) phase by defining a roto-displacive path in the configuration space. In our work, we take the Van der Waals interactions into account by DFT-D3 correction method [79-80] as implemented in the VASP software. The correlated nature of Mn $3d$ state are included by Hubbard-like corrections with repulsion energy $U$=3 eV and Hund coupling energy $J$=1 eV. Our calculations show that small variation of U and J does not change the main results of our study. In order to calculate the spin textures, the mean values of the sigma matrices are evaluated at the relevant electronic states with different $k$-points around a reference point in the Brillouin zone.

## Acknowledgments

This work is supported by NSFC 11825403, the Special Funds for Major State Basic Research (Grant No. 2015CB921700), the Program for Professor of Special Appointment (Eastern Scholar), the Qing Nian Ba Jian Program, and the Fok Ying Tung Education Foundation. F. Lou thanks Dr. K. Liu and J. Li for useful discussion. T. Gu thanks the kind hospitality of CNR-SPIN c/o Department of Chemical and Physical Science of University of L'Aquila where this project was initiated during the visiting period from 8$^{th}$ November 2017 to 14$^{th}$ January 2018. A.S. would like to thank the warm hospitality of W. Li, X.-H. Bu (Nankai University), H. Wu (Fudan University), W. Ren (Shanghai University) where this work was partially finalized.

## Author contributions

H.X. and A.S. proposed and supervised the project. F.L. and T.G. performed the first-principles calculations with the help from J.F., F.L. and T.G. prepared the initial draft of the paper. All authors contributed to the writing and revision of the paper. F.L. and T.G. contribute equally to this work.

## Data availability

The datasets generated and analyzed here are available from the corresponding author upon reasonable request.

## Competing financial interests

The authors declare no competing financial interests.

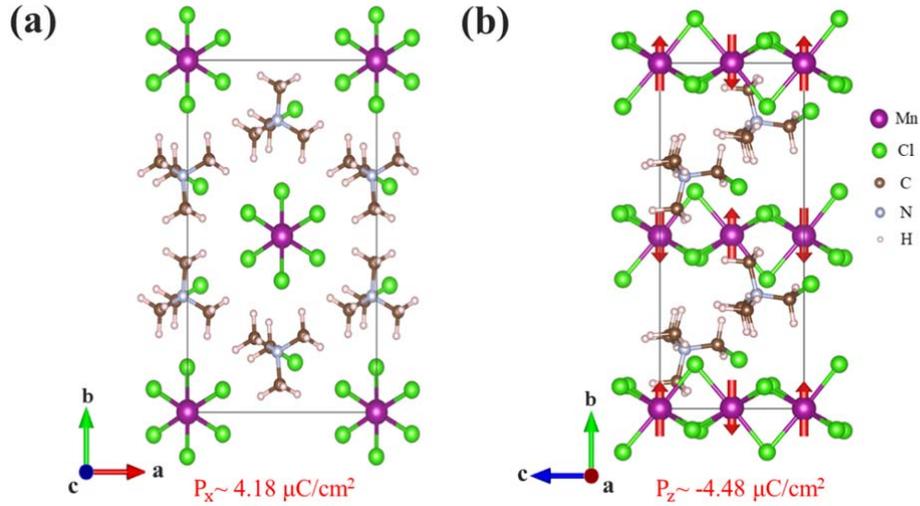

Fig 1. The atomic structures of TMCM-MnCl$_3$ in the $Cc$ phase from the top (a) and side (b) views, respectively. The magnetic configuration of G-type AFM$_y$ ($\mathbf{L_G}\sim y$) state is shown in (b). The AFM order parameter is defined as $\mathbf{L} = \sum_i \mathbf{S}^i - \sum_j \mathbf{S}^j$, where $\mathbf{S}^i$ ($\mathbf{S}^j$) is the spin moment along positive (negative) axis, respectively. Here, $\mathbf{L_G}\sim y$ denotes G-type AFM state with AFM order parameter (L) along $y$ [010] axis. The red arrows represent the local moment which is set to be along the $y$ direction. The polarization is along approximately [10$\bar{1}$] direction with P$_x$ ~ 4.18 μC/cm$^2$ and P$_z$ ~ 4.48 μC/cm$^2$.

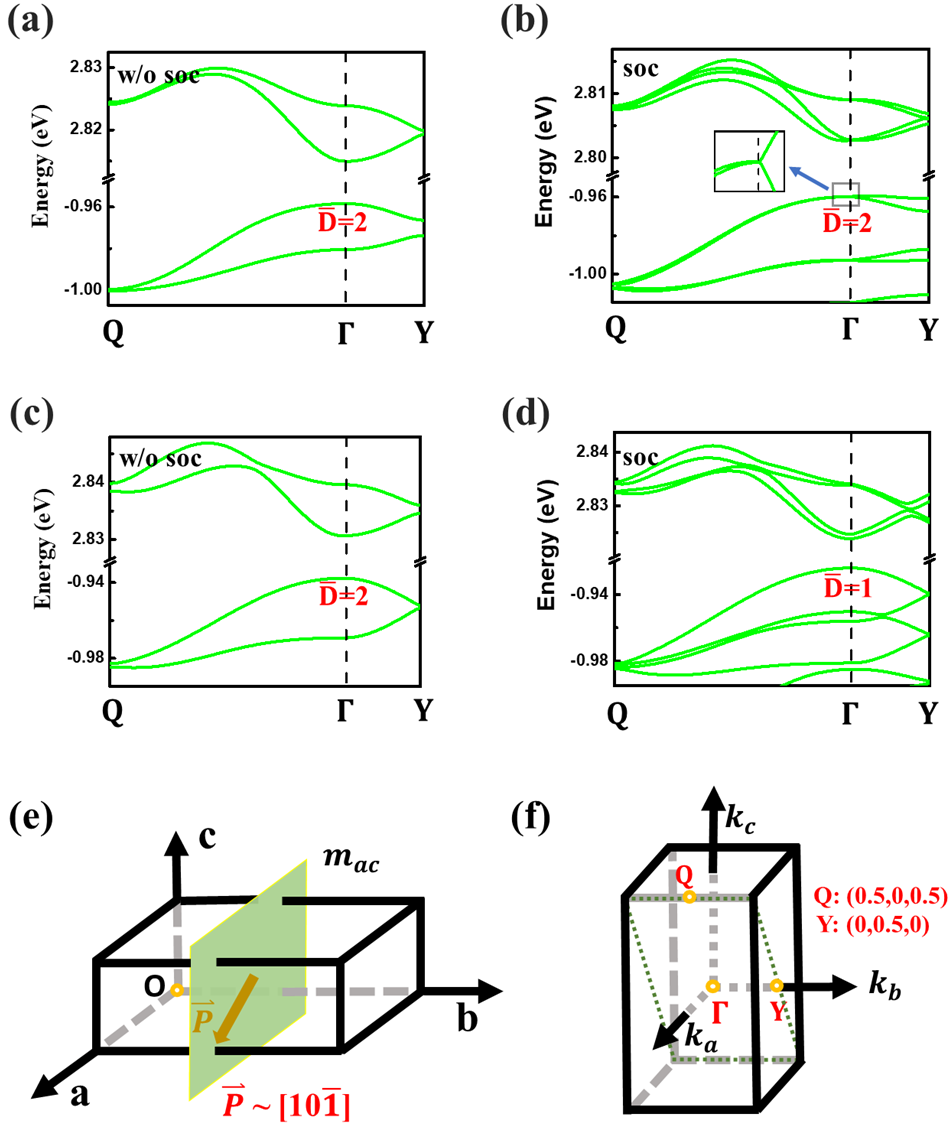

Fig 2. The band structures of G-type AFM (a,b) and C-type AFM (c,d) without SOC (a,c) and with SOC (b,d), respectively. In (b,d), the spin moment is set to be along the *y* direction. The abscissa is the Brillouin zone coordinate, and the ordinate is the energy, where the Fermi level is set to 0 eV. The magnified inset in (b) shows the SOC-induced band splitting around Γ point. The band degeneracy at Γ point is represented by $\bar{D}$. The crystal lattice accompanied with the mirror reflection are shown in (e) with the polarization along approximately $[10\bar{1}]$ direction. In (f), the first Brillouin zone with the symmetry path in band structure calculations. The olive-green section containing $k_b$ and $k_{ac}$ is adopted to draw the spin texture, where $k_b$ and $k_{ac}$ denote the $k$ path from Γ (0,0,0) to Y (0,0.5,0) and Q (0.5,0,0.5), respectively.

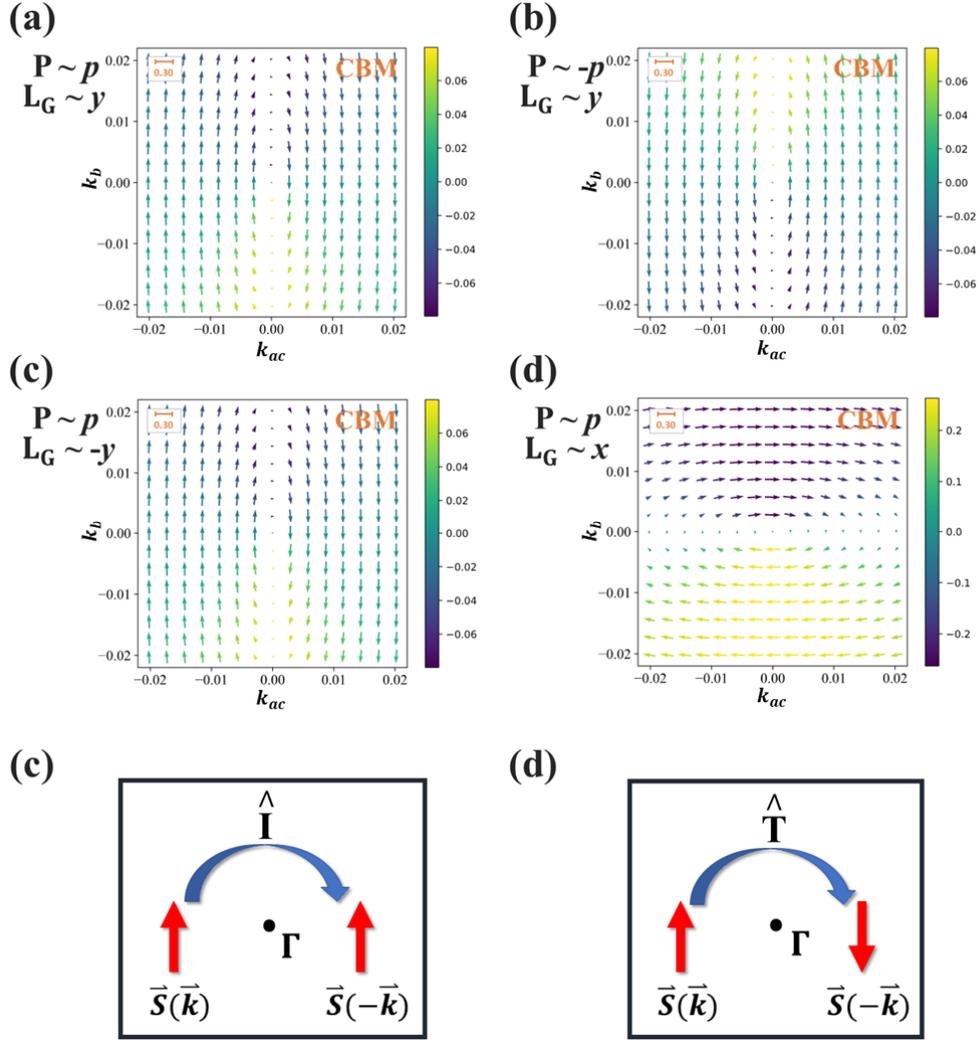

Fig 3. The interplay between ferroelectric ordering, magnetic ordering, and spin textures in G-type AFM state. Here we show the spin textures at CBM. **L$_G$~y** (**L$_G$~-y**) present G-type AFM state with spin moment along $y$ (-$y$) axis, respectively. **L$_G$~x** presents G-type AFM state with spin moment along $x$ axis. **P~p** (**P~-p**) denotes polarization along [10$\bar{1}$] ([$\bar{1}$01]), respectively. $k_b$ and $k_{ac}$ denote the $k$ path with the unit of Å$^{-1}$. The arrows refer to the in-plane orientation of spin and the scale is shown in the top left corner, while colors indicate the out-of-plane component. In (b), the polarization is switched to –$p$. In (c), the magnetic ordering is switched to -$y$ direction. In (d), the magnetic ordering is switched to $x$ direction. In (e) and (f), we show the schematic diagrams of spin transformation. In (e), the spin transformation under spatial inversion operator $\hat{I}$ (*i.e.* the reversal of polarization). In (f), the spin transformation under time reversal operator $\hat{T}$ (*i.e.* the reversal of AFM order parameter).

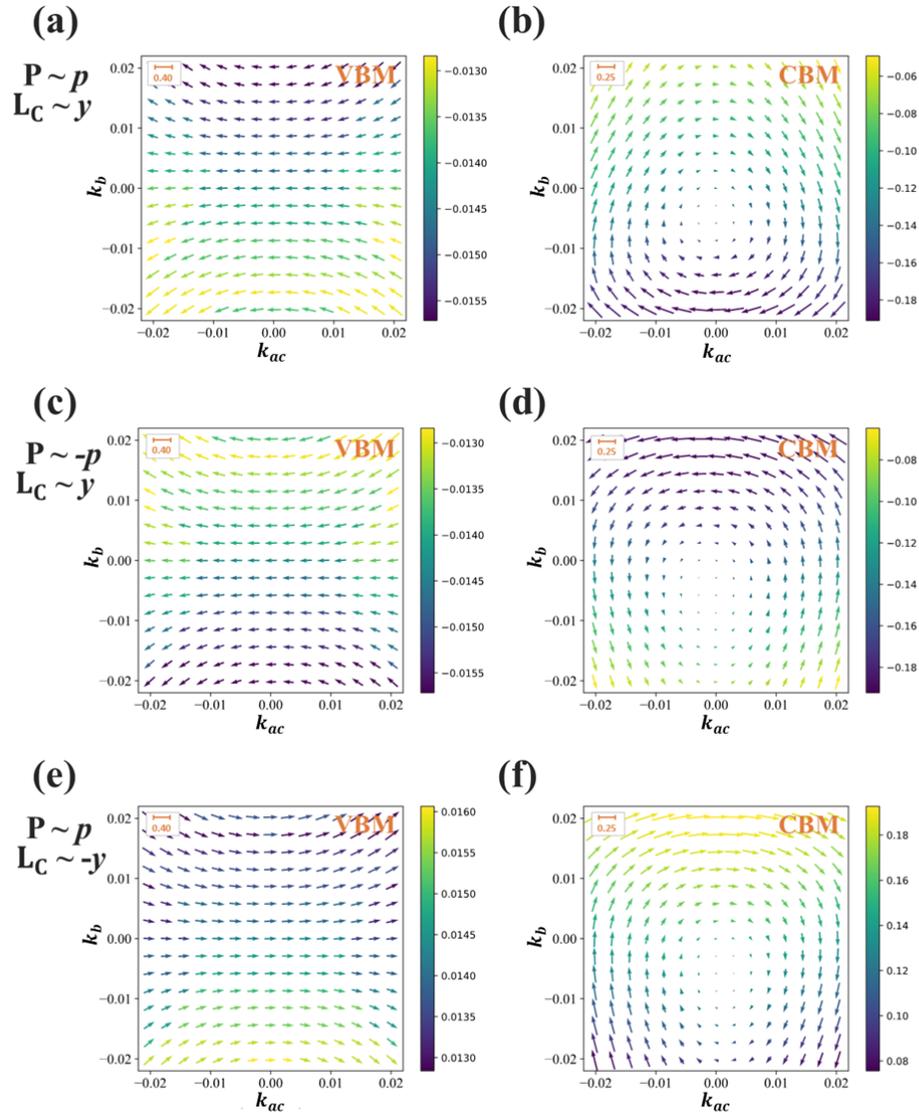

Fig 4. The interplay between ferroelectric ordering, magnetic ordering, and spin textures in C-type AFM state. Here we show the spin textures at VBM and CBM. **L$_C$**~*y* (**L$_C$**~*-y*) presents C-type AFM state with spin moment along *y* (*-y*) axis, respectively. In (c) and (d), the polarization is reversed from *p* to *–p*. In (e) and (f), the magnetic ordering is reversed from *y* to *-y* direction.